\documentclass[prl,aps,twocolumn,reprint,floatfix,nofootinbib]{revtex4-1}
\usepackage{amssymb,graphicx}
\usepackage{epsfig}
\usepackage[usenames]{color}

\newcommand{\beqn}{\begin{eqnarray}}
\newcommand{\eeqn}{\end{eqnarray}}
\newcommand{\beq}{\begin{equation}}
\newcommand{\eeq}{\end{equation}}

\begin{document}

\title{Ultrarelativistic Black Hole Formation}
\author{William E.\ East and Frans Pretorius}
\affiliation{
Department of Physics, Princeton University, Princeton, New Jersey 08544, USA 
}

\begin{abstract}
We study the head-on collision of fluid particles well within the kinetic energy dominated
regime ($\gamma=8$ to 12) by numerically solving the Einstein-hydrodynamic equations.
We find that the threshold for black hole formation is lower (by a factor of a few) than 
simple hoop conjecture estimates, and, moreover, near this threshold
two distinct apparent horizons first form postcollision and then merge.
We argue that this can be understood in terms of a gravitational focusing effect.
The gravitational radiation reaches luminosities of $0.014$ $c^5/G$, carrying $16\pm2\%$ 
of the total energy.
  
\end{abstract}

\maketitle

{\em Introduction.}---%
An important topic in high energy physics that remains poorly understood is the dynamics and outcome
of super-Planck scale particle collisions.
According to general relativity, kinetic energy, like all forms of energy, gravitates.  
This implies that at sufficiently high center of mass energies $E$, the gravitational force will eventually dominate
any interaction. Suppose one can localize the particles' wave functions at the moment of interaction
to be within a sphere of radius $R$; then, according to Thorne's hoop 
conjecture~\cite{thorne_hoop} (see also~\cite{1987PhLB..198...61T,1999hep.th....6038B,2002JHEP...06..057K}), 
if $R$ is less than the corresponding Schwarzschild radius $R_s=2 G E/c^2$, the gravitational interaction
will be so strong that a black hole (BH) will form. For particles satisfying the de Broglie relationship
the threshold for BH formation occurs at Planck energies.
There has been much interest in the past decade over the possible relevance of this 
to proton collisions at the Large Hadron Collider~\cite{2001PhRvL..87p1602D,2002PhRvD..65e6010G}
and cosmic ray collisions with the Earth's atmosphere~\cite{PhysRevLett.88.021303}, spurred
by theories of quantum gravity with small or warped extra 
dimensions~\cite{ArkaniHamed:1998rs,Antoniadis:1998ig,Randall:1999ee}
that present the possibility of a true Planck scale within reach of these processes.
To date no evidence for BH formation 
has been found~\cite{Heros:2007hy,Chatrchyan:2012taa}, though since the theories do not
make firm predictions for what the true Planck scale is, the high energy scattering
problem is worthy (beyond intrinsic theoretical interest) of further study.

Here we explore the purely classical gravitational properties of head-on ultrarelativistic collisions
(in four-dimensional asymptotically flat spacetime). This
ostensibly gives the leading order description of the process for energies sufficiently
above the Planck scale, as all nongravitational interactions will be hidden behind
the event horizon, implying that the particular model for the particles is irrelevant.
However, part of the motivation for this
study is to test this notion, and begin to investigate how it breaks down approaching
the threshold of BH formation (though again only at the classical level).

There have been several studies of ultrarelativistic collisions using BHs as model particles. 
Penrose~\cite{Penrose} first considered the head-on
collision of two Aichelburg-Sexl metrics~\cite{Aichelburg:1970dh}, each representing the
boost $\gamma\rightarrow\infty$ limit of the Schwarzschild metric (letting the mass $M$ go to zero such that
the energy $E=\gamma M$ is fixed, and note throughout we use geometric units $G=c=1$).
Though the spacetime to the causal future is unknown, 
a trapped surface is present at the moment of collision, giving an upper bound 
of $29\%$ for the radiated energy. Perturbative methods~\cite{D'Eath:1976ri,eath_payne}
allowed a direct calculation, estimating $16.4\%$ energy emitted. In~\cite{sperhake}, head-on
collisions up to $\gamma\approx 3$ were studied using numerical solutions of the
field equations; extrapolating the results to $\gamma\rightarrow\infty$ gave 
a value of $14\pm3\%$.
We briefly mention that studies of BH collisions for general impact parameters
using the  
trapped surface method for infinite boosts~\cite{Eardley:2002re}, and 
numerical simulations of finite boosts~\cite{Shibata:2008rq,Sperhake:2009jz} show that
considerably more energy can be radiated then.

However, as detailed in~\cite{uvbs}, the application of the infinite boost results to the
collision of massive particles at ultrarelativistic but subluminal speeds is not entirely clear. 
In this limit, the spacetime loses asymptotic flatness while the non-Minkowski part of
the spacetime becomes a two-dimensional shockwave. 
Moreover, BH collisions at any speed will necessarily produce a larger BH for sufficiently small impact parameter, 
and are not suitable for studying the threshold or dynamics of BH formation,
nor whether BH formation is the generic outcome regardless of the nature or compactness of the
colliding particles. Trapped surface calculations, as 
in~\cite{Eardley:2002re}, can be used to infer the dependence of BH formation
on impact parameter (which we do not consider here); however, they do not provide information
on the spacetime dynamics postcollision.
In~\cite{uvbs} a first attempt to address some of these
questions was made, where the ultrarelativistic collision of boson stars 
(solitons of a minimally coupled complex scalar field) was
studied numerically.  It was found
for boson stars with compactness $2 M/r\approx 1/20$ 
that a BH forms for boosts greater than $\gamma\approx2.9$, roughly one-third the
value $\gamma_{\rm h}=10$ predicted by applying the hoop conjecture at the time of collision. 
Whether the threshold is generically such a factor smaller than the hoop conjecture
estimate was unclear, first because only a single matter model was 
considered, but also because, though for $\gamma=2.9$ there is almost twice as much kinetic as rest mass
energy in the spacetime, this may not be high enough for the matter dynamics to be
irrelevant. Furthermore, due to difficulties disentangling gauge from gravitational wave (GW)
dynamics, no estimates of the radiated energy were made.

In this Letter we also study black hole formation in head-on particle collisions.
However, we use  
perfect fluid ``stars'' as the model particles. 
To begin with, this allows us to further test the generality of
the above arguments in a case where gravity  
would be opposed by the tendency of the fluid to become highly pressurized on 
collision and disperse.  Second, the nature of fluid stars, not having 
small-scale internal oscillations as boson stars, as well as a new method for constructing initial
data~\cite{idsolve_paper}, permits us to explore significantly higher boost collisions
where the ratio of kinetic to rest mass energy is of order 10:1.
An independent work with the same matter model used here was recently presented in~\cite{Rezzolla:2012nr},
though as with~\cite{uvbs} it focuses on regimes where this ratio is at most $\approx$ 2:1.

We find that BHs are formed above a critical boost $\gamma_c$ that is a factor of a few less
than the hoop conjecture estimate.
A new phenomenon we present here is, for boosts slightly above $\gamma_c$, we 
observe {\em two} separate apparent horizons (AHs) form shortly {\em after} the collision, which some time later
are encompassed by a single horizon that rings down to a Schwarzschild BH. 
We argue that this can be qualitatively understood as due to the strong focusing
of the fluid elements of one star by the boosted spacetime of the other,
and vice versa, using a geodesic model similar to that in~\cite{Kaloper:2007pb} for BH formation
in the scattering problem.
We also study the GWs emitted in this regime 
for the first time and find that for the $\gamma=10$ BH forming case $16\pm2\%$ of the energy of the 
spacetime is radiated (the extrapolation described in~\cite{sperhake} suggests this should be $94\%$ of the $\gamma=\infty$ limit).
For subthreshold cases, the strong focusing leads to high fluid pressures
that cause the stars to explode outward. 
In what follows, we outline the equations we are solving, the numerical methods for doing so,
and the setup of the initial data.  We then present the results of our simulations, compare
them to geodesic focusing, and end with concluding remarks.

{\em Methodology.}---%
We numerically solve the Einstein field equations, in the generalized harmonic formulation, 
coupled to a perfect fluid using the code described in~\cite{code_paper}.
For simplicity we use the $\Gamma=2$ equation of state. 
We use a variation of the damped harmonic gauge~\cite{uvbs,dampedGauge}
that corresponds to equation (A15) in~\cite{dampedGauge} with $p=1/4$. 

We take advantage of the axisymmetry of a head-on collision to reduce the numerical
grid to two dimensions   
and use seven levels of mesh refinement
where the finest level covers the equatorial and polar radii of the star by approximately 830 and (due
to Lorentz contraction) $830/\gamma$ points, respectively.
For the $\gamma=10$ case, to estimate truncation error we also ran simulations with 1.5 and 2 times the resolution.
Unless otherwise stated, results from this case are from the high resolution run. 

Initial data are constructed using free data from two identical, boosted solutions of the
Tolman-Oppenheimer-Volkoff equations with a polytropic condition, and then solving the 
constraint equations in the conformal thin-sandwich formulation as described in~\cite{idsolve_paper}.
With this method, the ``spurious" gravitational radiation is much 
smaller than the physical signal (see Fig.~\ref{gw_fig} below). 
We choose isolated star solutions with compaction $2 M_*/R_*=1/40$, where
$M_*$ and $R_*$ are the gravitational mass and radius, respectively, of the star in its rest frame.
They are boosted towards 
each other with Lorentz factor $\gamma$, 
at an initial separation of $d=534M_*$.  We consider cases with $\gamma=8$, 8.5, 9, 9.5, 10, and 12,
though most of our detailed results are from $\gamma=8$ and $10$.

\begin{figure*}
\begin{center}
\includegraphics[width=1.35in,draft=false]{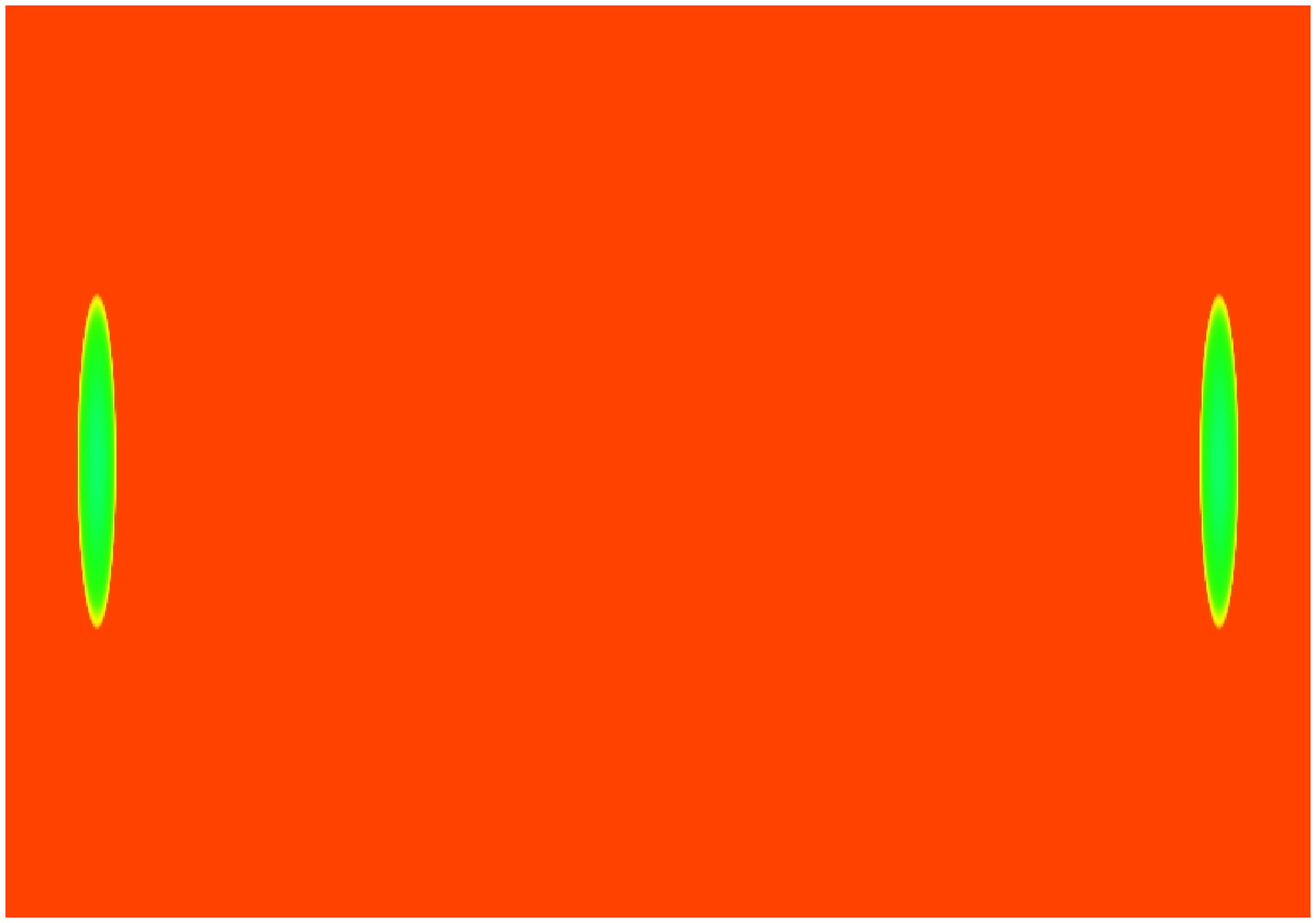}
\includegraphics[width=1.35in,draft=false]{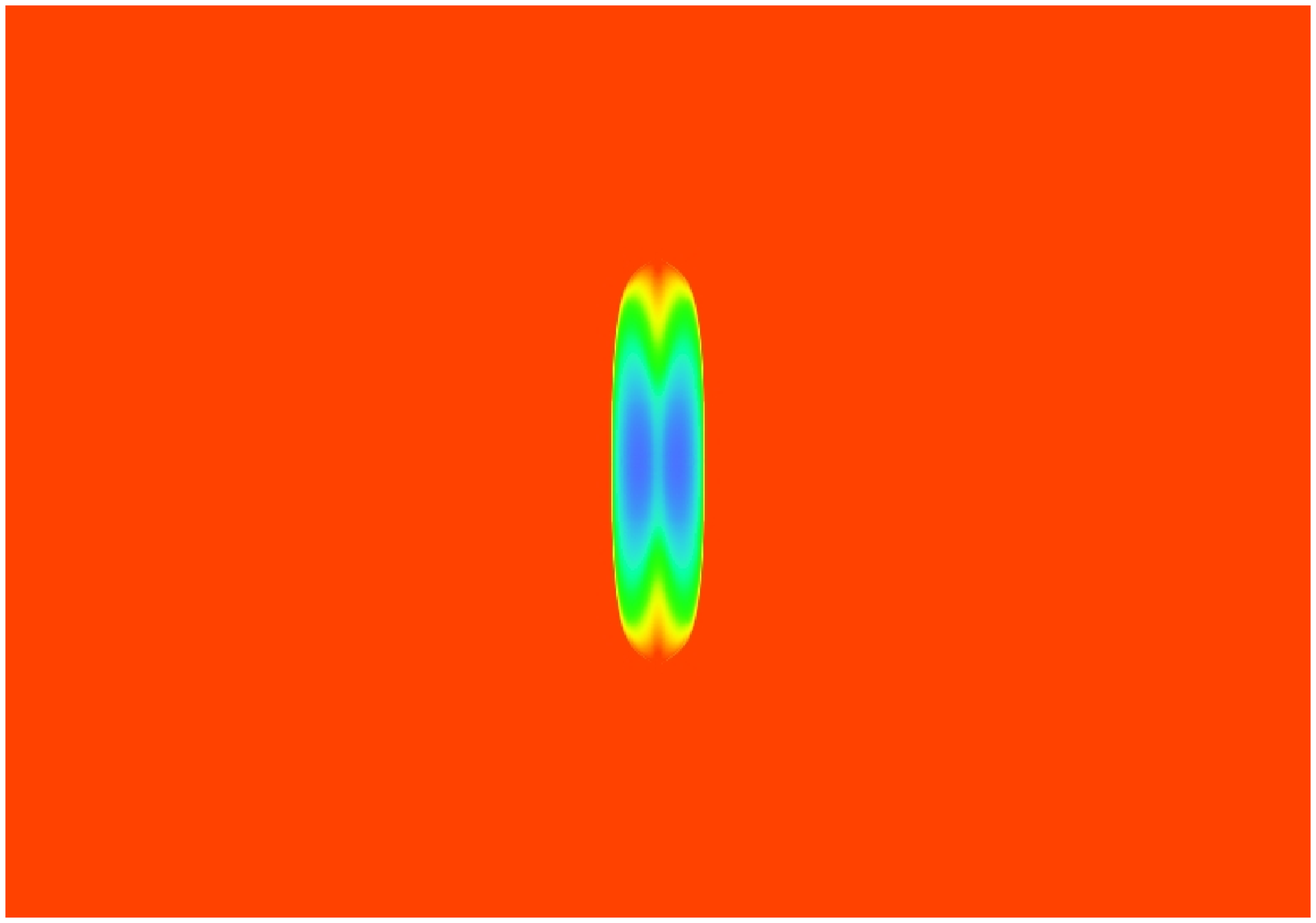}
\includegraphics[width=1.35in,draft=false]{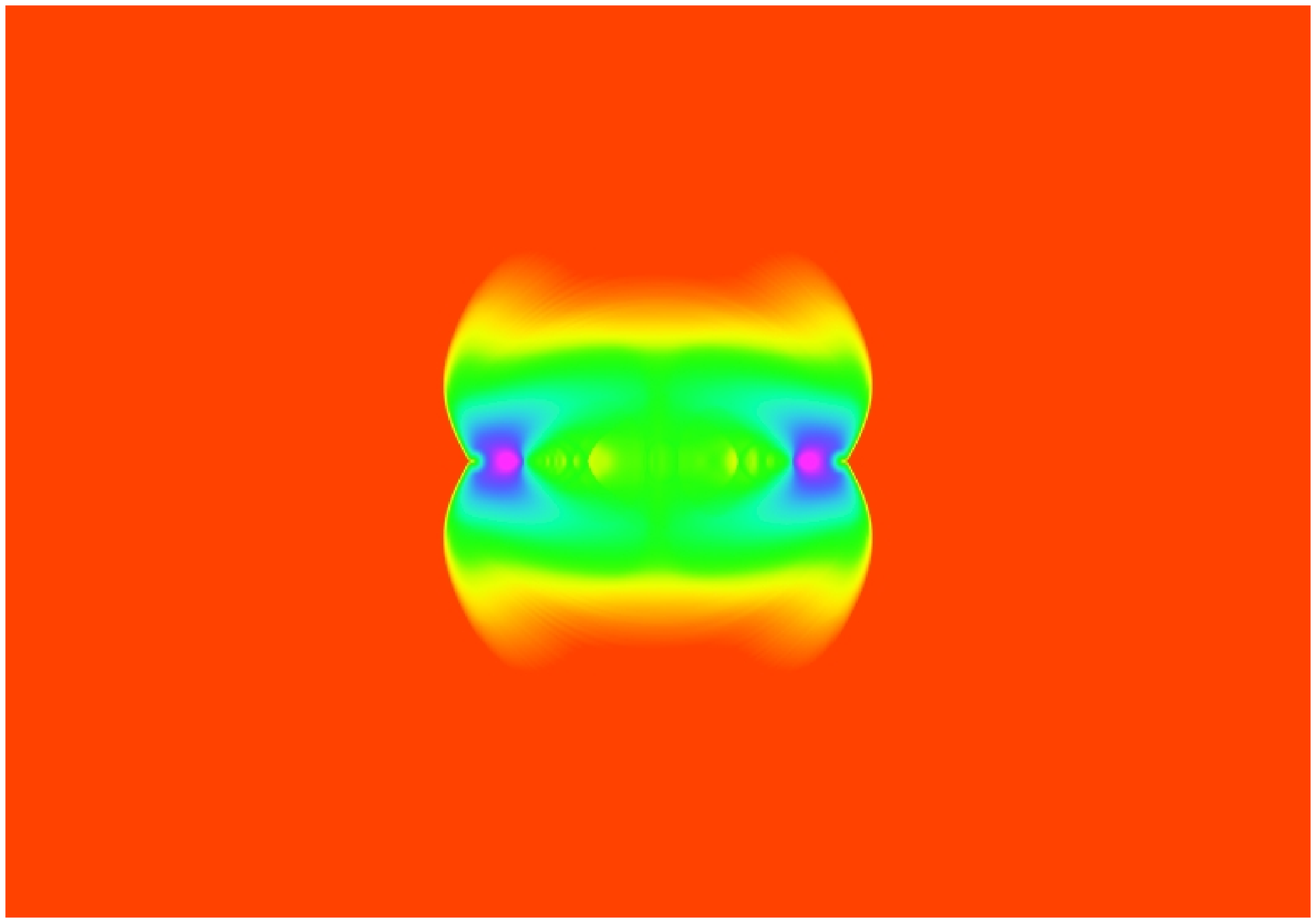}
\includegraphics[width=1.35in,draft=false]{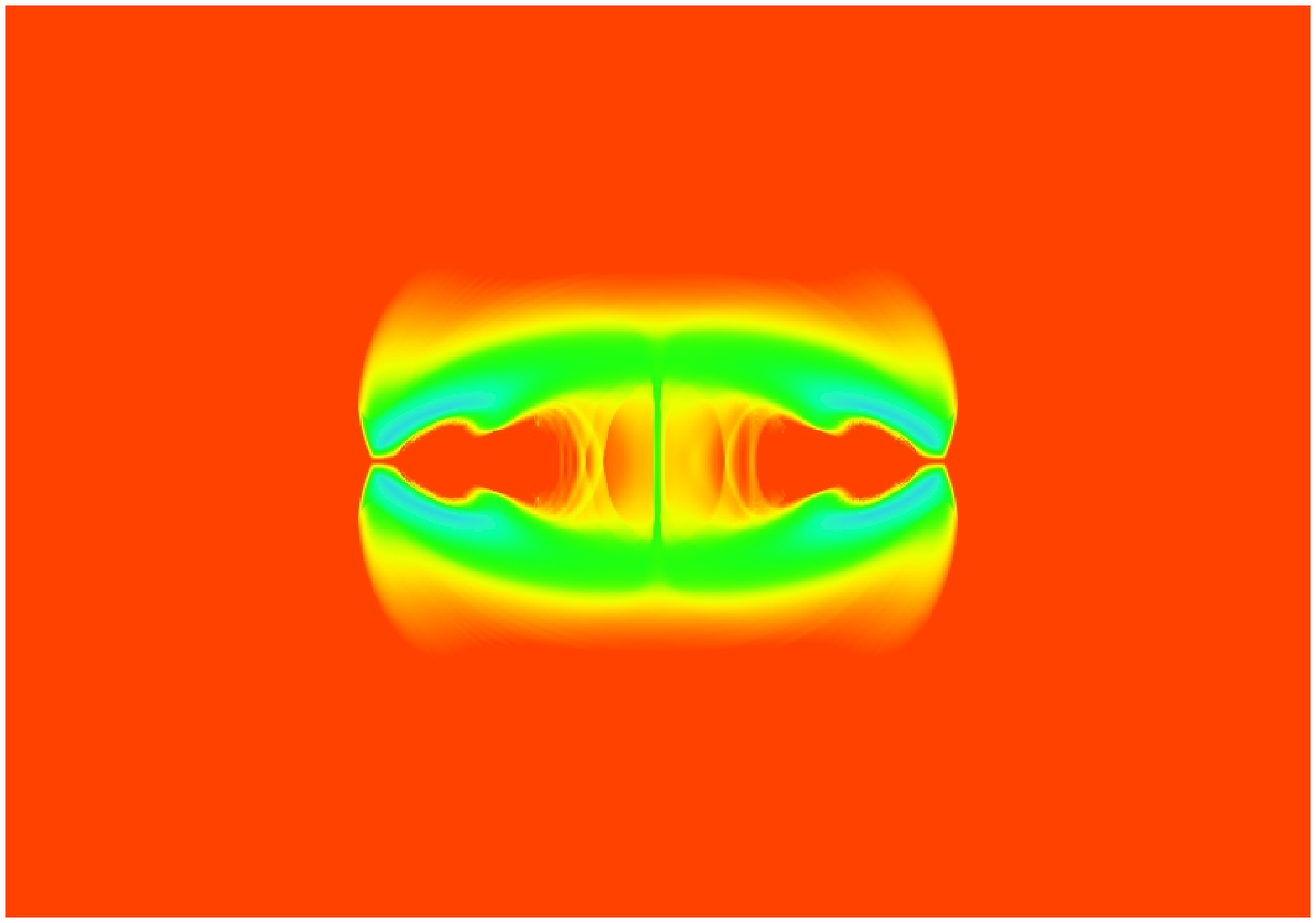}
\includegraphics[width=1.35in,draft=false]{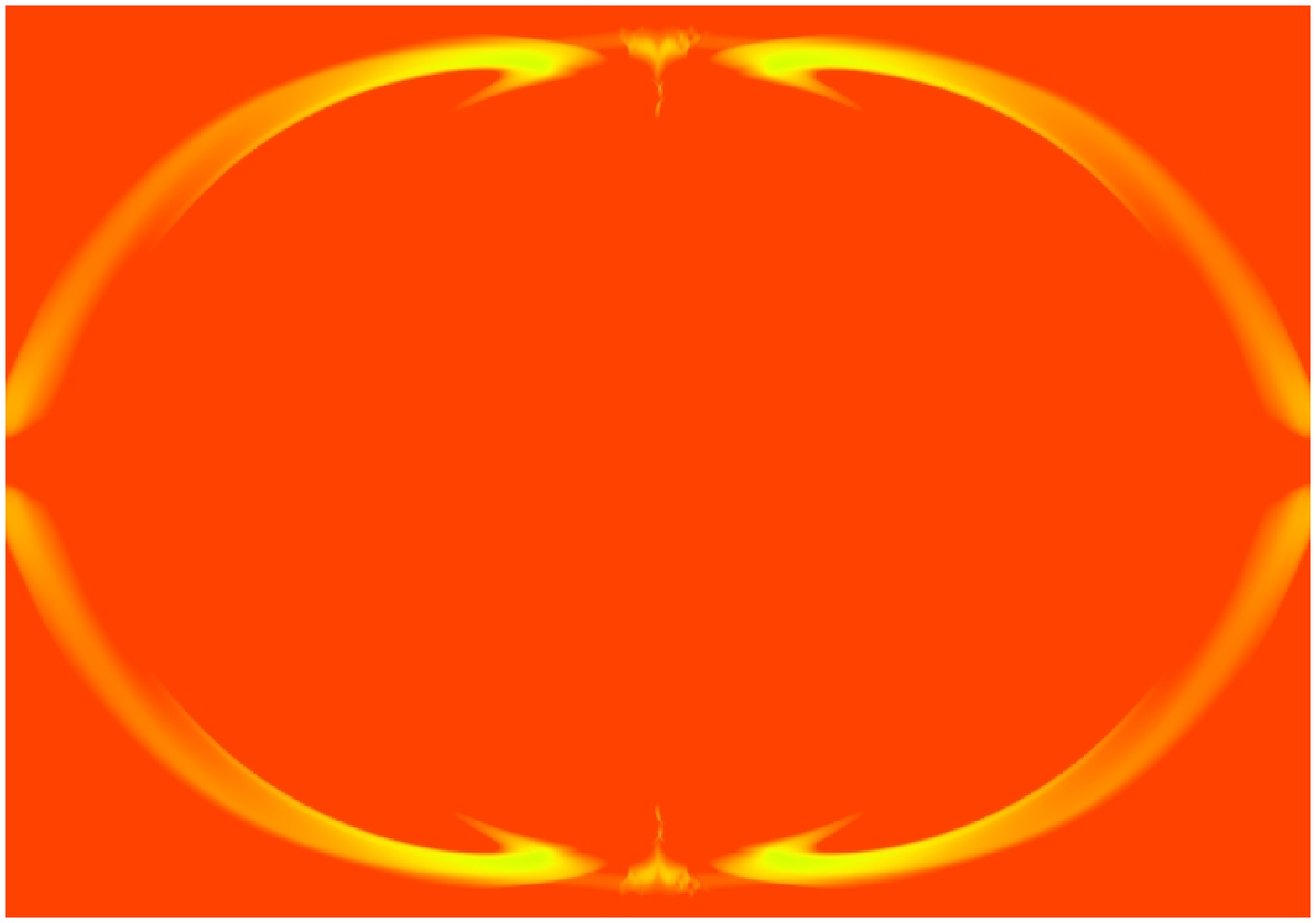}
\\
\vspace{0.02 in}
\includegraphics[width=1.35in,draft=false]{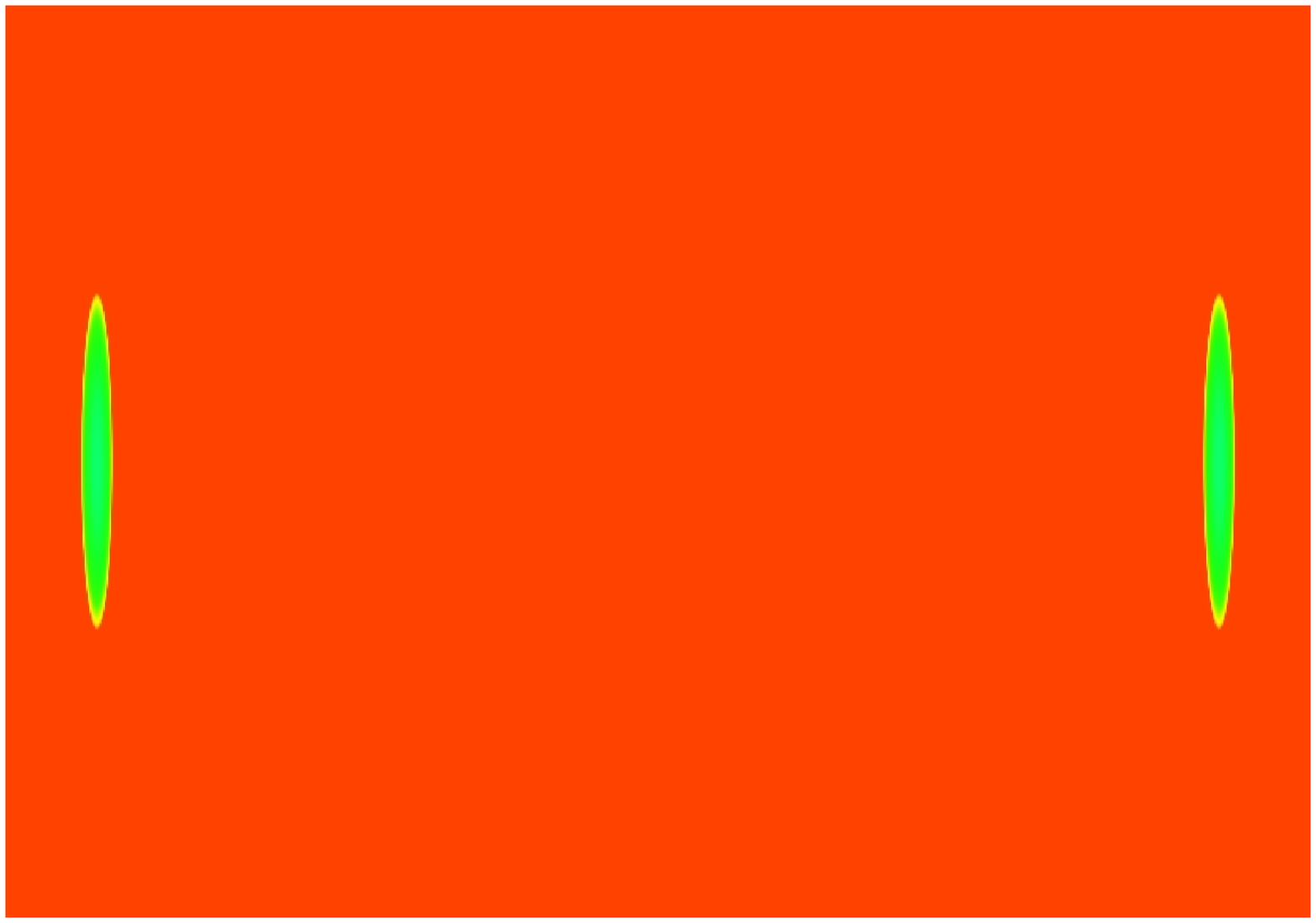}
\includegraphics[width=1.35in,draft=false]{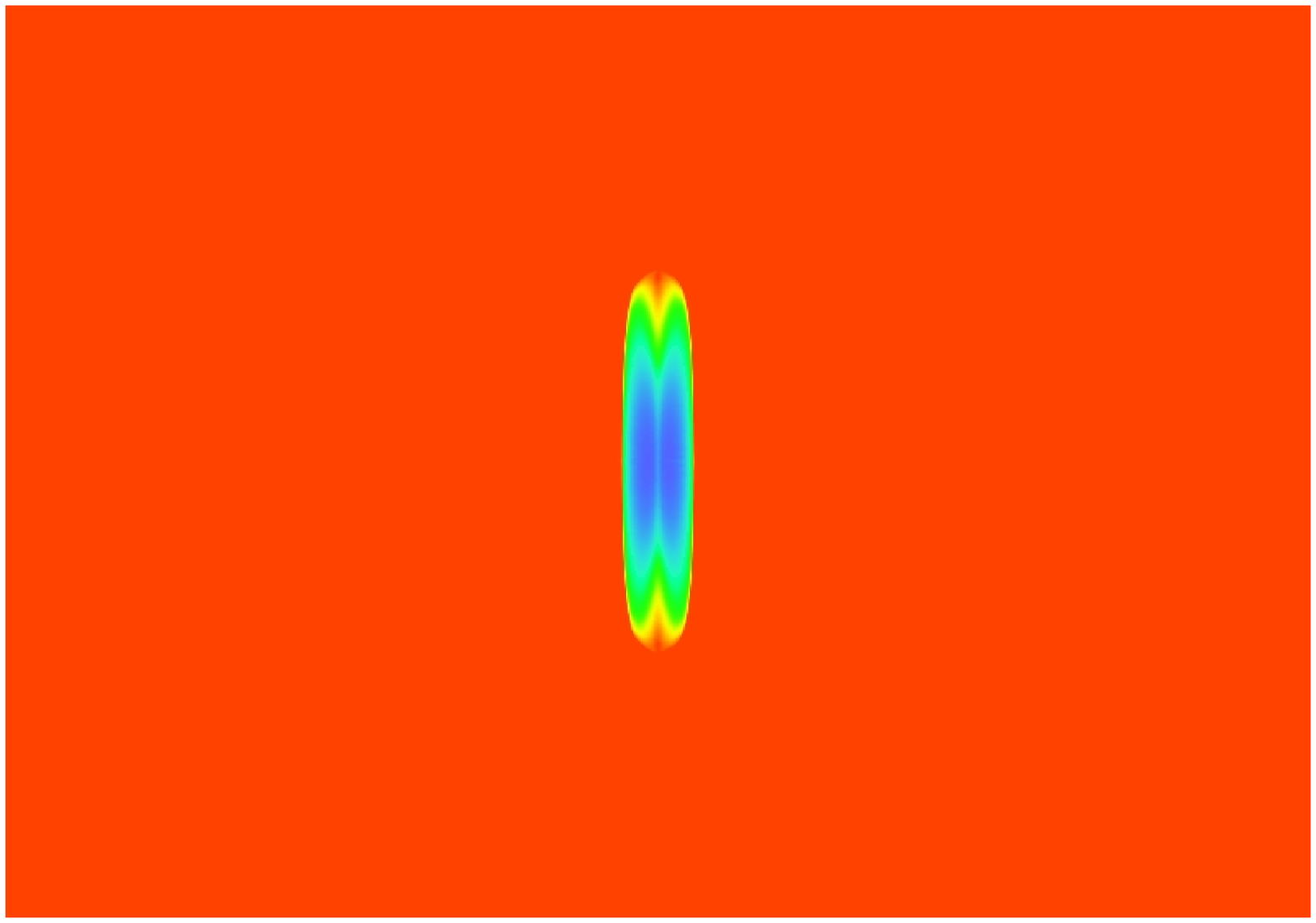}
\includegraphics[width=1.35in,draft=false]{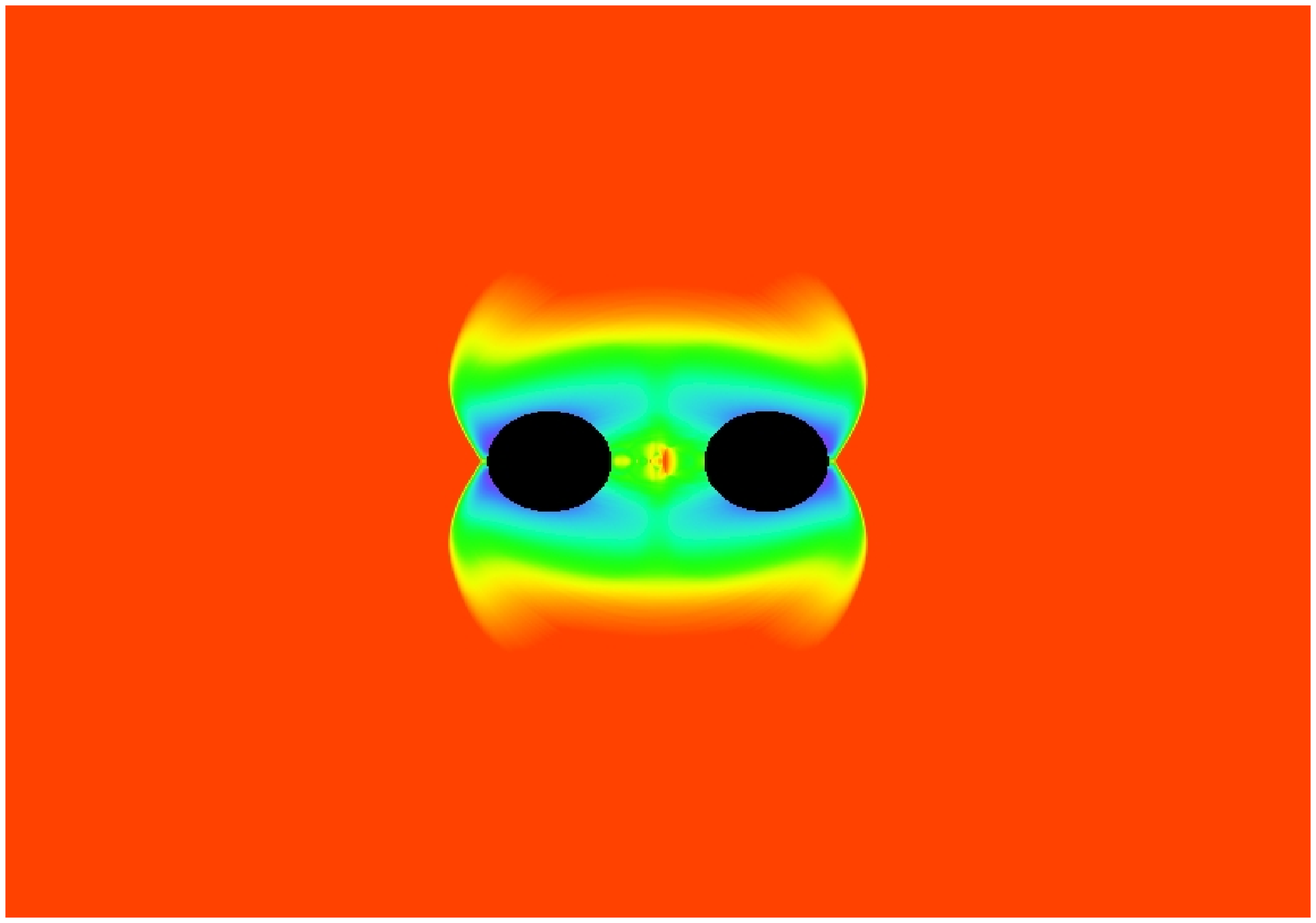}
\includegraphics[width=1.35in,draft=false]{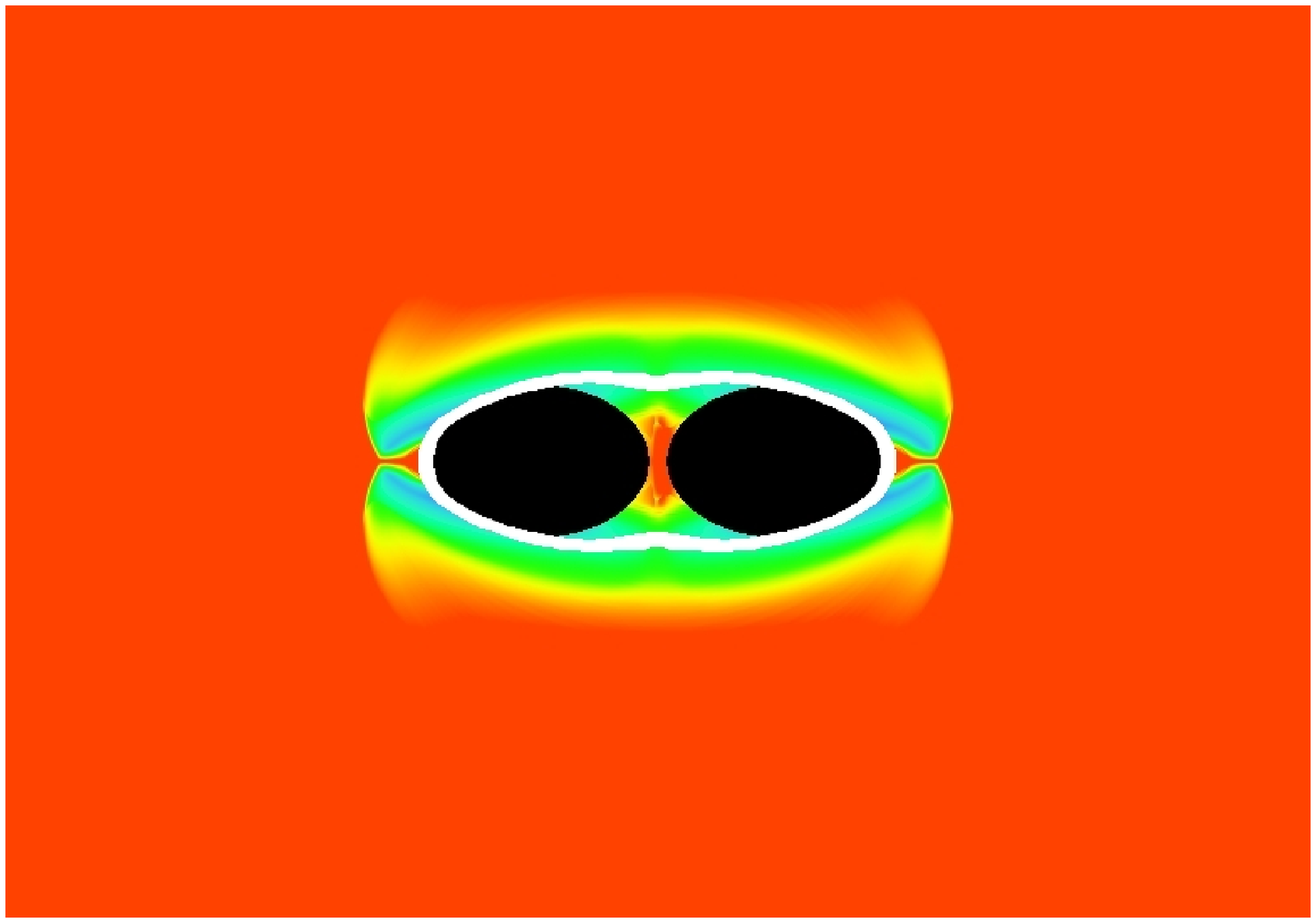}
\includegraphics[width=1.35in,draft=false]{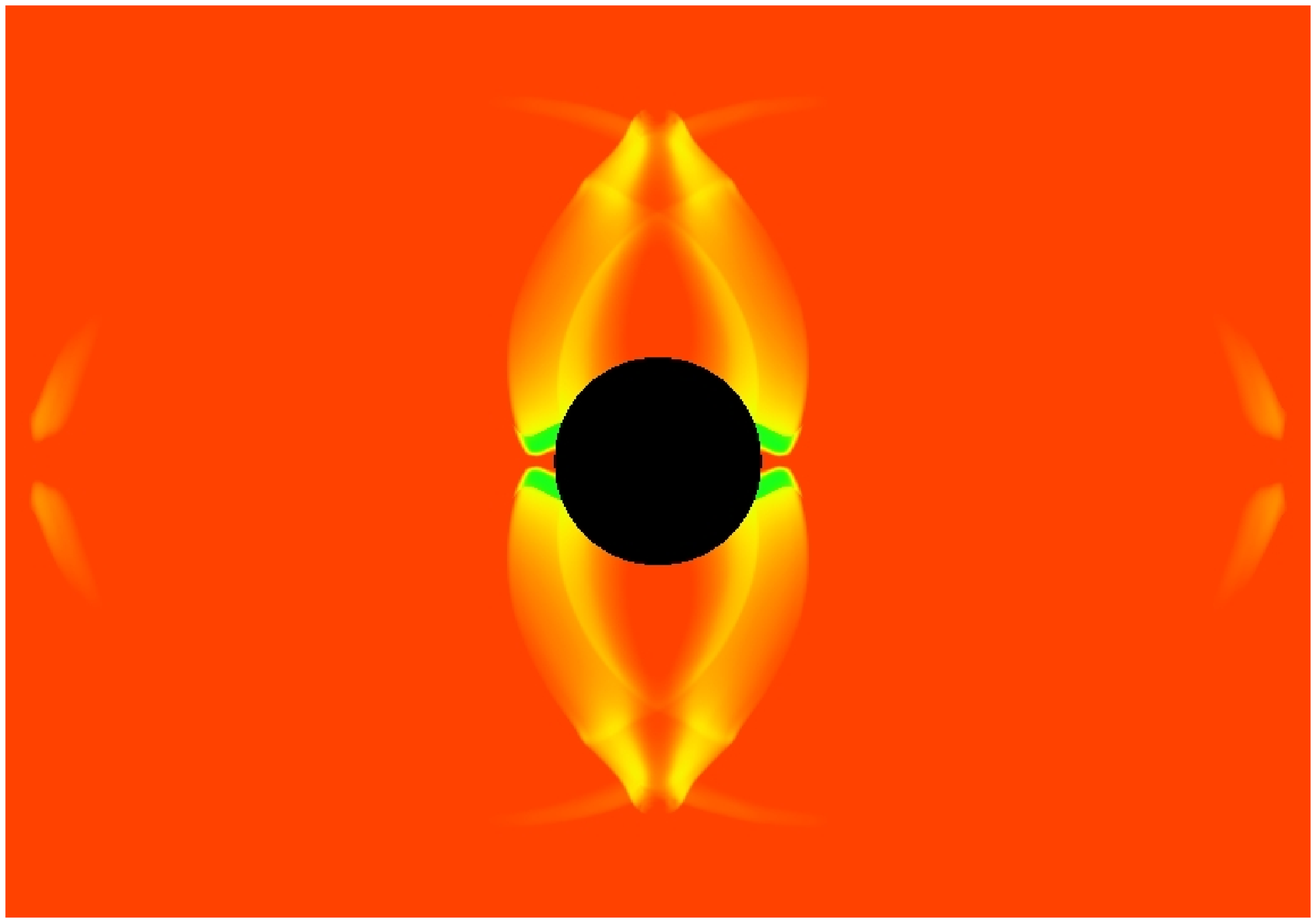}
\end{center}
\caption{
Snapshots of rest mass density on a logarithmic scale from $10^{-2}$ to $10^2$ times the initial maximum density,
for simulations with $\gamma=8$ (top) and $\gamma=10$ (bottom)
at times (left to right) $t=0$, the initial time; $t=300M_{*}$, shortly {\em after} collision; 
$t=375M_*$, after the appearance of the smaller AHs in the $\gamma=10$ case; 
$t=424M_*$, after the appearance of the third, encompassing AH (white outline) in the $\gamma=10$ case;
and $t=700M_*$.
The black regions are best-fit ellipses to the AHs. 
\label{snapshots}
}
\end{figure*}

{\em Results.}---%
We find that BH formation {\em does} occur in the ultrarelativistic collision of
fluid particles with the aforementioned compaction for boost factors $\gamma \geq 8.5 \pm 0.5$ (the
uncertainty is from the sampling resolution of our survey in $\gamma$).
This is $\sim 2.4$ times smaller than the hoop conjecture threshold of $\gamma_{\rm h}\approx 20$.
In Fig.~\ref{snapshots}, we show snapshots of the rest mass density for a subcritical case with $\gamma=8$
and for a supercritical case with $\gamma=10$.  In the former, after the collision, the matter focuses 
down into two high density regions which then explode outward. 
In the latter, instead of exploding,
two identical AHs appear surrounding these regions. (It should be noted that the existence of the 
initially disjoint AHs does not preclude the possibility of a single encompassing event horizon.) 
The AHs then fall towards each other with a third,
encompassing AH appearing afterwards. 

\begin{figure}
\begin{center}
\includegraphics[width=3.2in,draft=false]{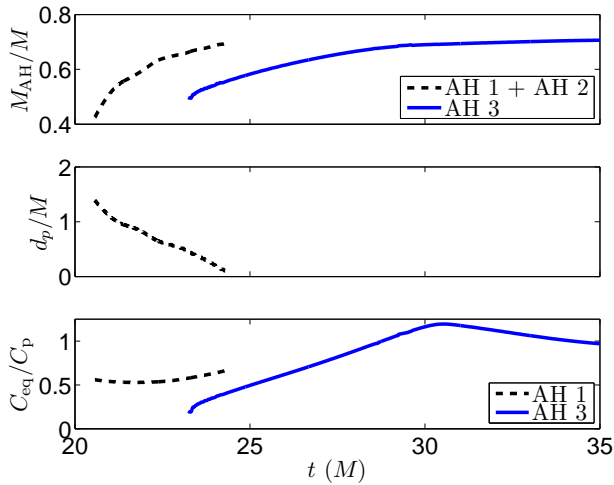}
\end{center}
\caption{
Apparent horizon properties for the $\gamma=10$ case. 
The AHs labeled 1 and 2 are the two identical (when mirrored about the collision
plane) ones that appear first and are later encompassed by AH 3.
We show (top to bottom): the irreducible masses, the proper distance $d_p$
between (and exterior to) AHs 1 and 2 measured along the collision axis, 
and the ratio of the equatorial to polar circumferences. 
\label{ah_fig}
}
\end{figure}

In Fig.~\ref{ah_fig}, we show the irreducible mass of the AHs, proper distance between the smaller AHs,
and the ratio of the proper equatorial and polar circumferences $C_{\rm eq}/C_{\rm p}$ for $\gamma=10$.  
The two smaller AHs are born rather prolate with $C_{\rm eq}/C_{\rm p}\sim0.6$.
Together they have mass $> 0.4M$ where $M\approx 2\gamma M_*$ is the total spacetime mass; 
i.e., they contain a significant amount of what was originally kinetic energy.  
When the third encompassing AH appears it initially has less irreducible mass (though greater area) than the sum 
of the smaller AHs.  It is also extremely distorted with $C_{\rm eq}/C_{\rm p}\sim 0.2$
and an equatorial circumference that is less than the smaller AHs, suggesting more of a dumbbell shape.   
This contrasts with what is found in ultrarelativistic black hole collisions where 
$C_{\rm eq}/C_{\rm p}\sim 1.5$ initially~\cite{sperhake}, 
consistent with a disk shaped AH. 

In Fig.~\ref{gw_fig}, we show the GW power associated with
different spherical harmonics
for $\gamma=10$, and the early part of the GW power for $\gamma=8$.
(Because of the symmetries here only the even $l$, $m=0$ harmonics are nonzero.) 
For $\gamma=10$, $16\pm2\%$ of the initial spacetime energy is radiated as GWs,
with a peak luminosity of $0.0137\pm1\%$ (the error bars include
estimates of the truncation error and finite radius extraction effects).  
The mass of the final BH is $\approx 0.72 M$, suggesting the remaining $12\%$
of the energy is carried off by the $\approx 32\%$ of the initial rest mass that remains outside the final BH
by the end of the simulation.
Measuring the contributions to the total energy from higher $l$ modes relative to the $l=2$ component we get that 
$E_{4}/E_{2}=0.19\pm 0.01$,  $E_{6}/E_{2}=0.073\pm 0.001$, and $E_{8}/E_{2}=0.040\pm 0.002$. 
The substantial amount 
of energy in higher modes is consistent
with results from ultrarelativistic BH collisions. Also, the zero-frequency limit 
combined with an $l$-dependent frequency cutoff $\omega_c=l/(3\sqrt{3}M)$ set by BH 
quasinormal frequencies~\cite{PhysRevD.81.104048} predicts corresponding values of 
0.22, 0.09, and 0.05.
For $\gamma=8$ we can only extract the GW signal before
the fluid outflow crosses the extraction sphere. Before this time,
the GW signal looks qualitatively similar to the
$\gamma=10$ case  
and contains $10\%$ of the energy of the spacetime.

\begin{figure}
\begin{center}
\includegraphics[width=3.2in,clip=true,draft=false]{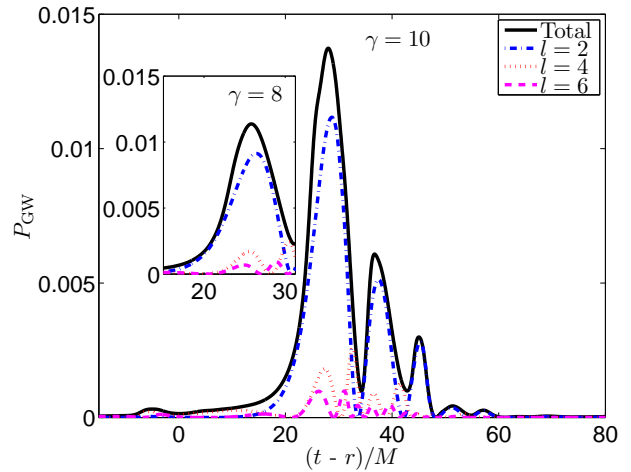}
\end{center}
\caption{
Total and spin-weight 2 spherical harmonic decomposition of power in GWs from the $\gamma=10$ case (with $\gamma=8$
in the inset).  The small feature at $(t-r)<0$ is an artifact of the initial data and is not included in the
estimate of energy.
\label{gw_fig}
}
\end{figure}

Cases with $\gamma=9.5$, 9.0, and 8.5 also first form two disjoint AHs with increasing initial separation, the smaller the boost. 
However, we were unable to follow these cases through merger before numerical instabilities
set in on the excision surface. The reason, we believe, is the smaller boosts form more distorted AH 
shapes, and our current approach of excising based on the best-fit ellipse to the AH shape is inadequate. 
We have also been unable to obtain robust results for significantly 
higher Lorentz factors due to high frequency numerical instabilities that develop at the surface of the 
boosted stars; 
however, it seems that the third AH appears at nearly the same time 
in this gauge as the first two AHs at
$\gamma \sim 12$ for the stars considered here, and for larger boosts we expect a
single AH to form at collision.

\begin{figure}
\begin{center}
\includegraphics[width=3.2in,draft=false]{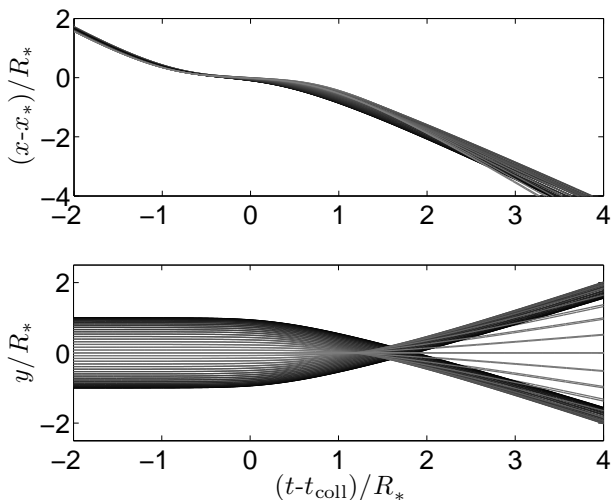}
\end{center}
\caption{
Focusing of a set of geodesics in a boosted star spacetime with $\gamma=10$ and $2 M_*/R_*=1/40$.  Shown are the $x$ coordinate relative
to the center of the boosted star which is at $x_*=vt$ (top) and the $y$ coordinate (perpendicular to the boost axis) of the geodesics as 
a function of coordinate time.   
\label{geodesic_fig}
}
\end{figure}

{\em Geodesic focusing.}---%
To illustrate the manner in which a boosted star may act like a gravitational lens and, during collision, focus the matter of the other
star, we consider a simplified scenario in a spacetime consisting of a single boosted star.  We
follow a set of geodesics coming from the opposite direction with the same Lorentz factor, initially 
distributed to fill out the volume of what would have been the other boosted star (i.e.,
we replace the second star with  
tracer particles).
These geodesics are shown in Fig.~\ref{geodesic_fig} for $\gamma=10$ with the same compaction
star  
described above. As these geodesics pass through the boosted star they become focused 
in the direction orthogonal to the boost axis while spreading out along the boost axis.  The greatest focusing 
(i.e., when the separation between the geodesics in either direction is smallest) occurs at a distance of 
$\approx1.5R_*$ away from the star and reduces the radius
by a factor of $\approx 4$ (roughly consistent, with the caveat of coordinate differences,
with 
the full problem---see Figs.~\ref{snapshots} and~\ref{ah_fig}).  
This contraction is sufficient to get to the BH formation threshold
if we assume that this focusing also converts sufficient translational energy to radial
inflow that it is valid to apply the hoop conjecture to the star's total energy
in this frame. Evidence for this assumption  
comes from the temporary slowdown of the 
translational velocity  
seen in Fig.~\ref{geodesic_fig} (though somewhat before maximum focusing),
and from the full simulations where in the $\gamma=10$ case
the two AHs move toward each other,  
and in the $\gamma=8$ case postcompression the fluid flow is largely radial. 
In the ultrarelativistic limit, this geodesic focusing factor is mainly a 
function of the ratio $\gamma M_*/R_*$, and similar
results are obtained for larger boosts with correspondingly less compact stars. 
This simplistic treatment of course ignores the effects of pressure and
nonlinear gravitational interactions.

{\em Conclusions.}---%
\label{conclusions}
In this Letter, we considered the head-on collision of self-gravitating fluid
stars in the regime where the ratio of kinetic to rest mass energy in the
spacetime is $\sim$10:1.
We find above a critical 
boost $\gamma_c=8.5\pm0.5$ that BHs do form. The dynamics of the solution,
and a simple geodesic model similar to~\cite{Kaloper:2007pb}, suggest that
near threshold the strong focusing nature of the spacetime sourced by one
boosted star on the other, and vice versa, causes the energy to be concentrated
postcollision around two focal points on axis. In the subcritical case, the material explodes outward from these
points, consistent with~\cite{uvbs,Rezzolla:2012nr}; however,
just supercritical we find two distinct AHs that initially form around
the focal points. 
This focusing also offers an intuitive explanation
for why the threshold in cases studied to date is systematically less
than hoop conjecture estimates (here $\gamma_c/\gamma_h\sim0.4$, 
with the boson star collisions $\gamma_c/\gamma_h\sim0.3$ for $\gamma_c\sim2.9$~\cite{uvbs},
and similar factors were found in~\cite{Amati:2007ak,Veneziano:2008zb,Marchesini:2008yh} for the scattering problem
using a perturbative  
model). 

For the $\gamma=10$ supercritical case, we find
$16\pm 2\%$ of the total energy is radiated gravitationally, consistent with results
extrapolated from $\gamma\approx3$ BH collisions~\cite{sperhake}, and 
perturbative calculations of the infinite boost limit~\cite{D'Eath:1976ri,eath_payne}.
Moreover, the leading order spherical-harmonic multipole structure of the waves 
is consistent with point-particle approximations and the BH case~\cite{PhysRevD.81.104048},
both super and subcritical, in the latter prior to obscuration of the
waveform by matter outflow.

This suggests three different regimes in the head-on collision
of ultrarelativistic, nonsingular model particles in general
relativity, for sources that have sufficiently {\em low} compactness such that $\gamma_c\gg 1$.
For $\gamma\ll \gamma_c$, gravity plays little role, and the dynamics
is governed by that of the matter; for $\gamma\gg\gamma_c$, we expect universal behavior; 
i.e., any particle model will give the same {\em quantitative} spacetime dynamics;
however, in the intermediate regime $\gamma\sim\gamma_c$ 
both gravitational and matter dynamics will be important.
Ignoring quantum effects and studying the nature of super-Planck scale particles
collisions using general relativity is arguably robust only 
when $\gamma\gg \gamma_c$,
though perhaps some insights can still be drawn from classical general relativity in the intermediate regime.

The intermediate regime includes the threshold of BH formation and corresponding,
{\em matter-dependent} critical phenomena~\cite{choptuik}.
We conjecture approaching $\gamma_c$ 
may generically result in two critical solutions 
unfolding postcollision
about the geodesic focal points of the two colliding particles
(we speculate the reason why two distinct AHs were not seen in~\cite{uvbs,Rezzolla:2012nr}
is the compactness is not sufficiently low to have $\gamma_c\gg 1$.)
For $\Gamma=2$ fluid stars, it would be interesting to see whether
the critical solution is the type I unstable starlike solution found for 
lower $\gamma_c$'s~\cite{Rezzolla:2012nr}, or the type II self-similar solution arising
in the kinetic energy dominated regime~\cite{Noble}.
It would also be interesting to explore collisions with nonzero
impact parameters. 
This would allow a better comparison to BH collisions, which
do not have a threshold for BH formation, but do have 
two distinct end states as a function of impact parameter:
a large BH or two unbound BHs.

We thank Emanuele Berti, Vitor Cardoso, Uli Sperhake, and Branson Stephens for useful conversations.
This research was supported by the NSF
Graduate Research Program under Grant No. DGE-0646086 (WE), NSF
Grant No. PHY-0745779, and the Alfred P. Sloan Foundation (FP).
Simulations were run on the Woodhen and Orbital clusters at Princeton University.

\bibliographystyle{h-physrev}
\bibliography{nsns}
\end{document}